# Layered semiconductor devices with water top-gates: High on-off ratio field-effect transistors and aqueous sensors

Yuan Huang,[1] Eli Sutter,[2] and Peter Sutter[3*]

[1]*Institute of Physics, Chinese Academy of Sciences, Beijing, 100190, China*
[2]*Department of Mechanical & Materials Engineering, University of Nebraska-Lincoln, Lincoln, NE 68588, USA*
[3]*Department of Electrical & Computer Engineering, University of Nebraska-Lincoln, Lincoln, NE 68588, USA*

## Abstract

Layered semiconductors show promise as channel materials for field-effect transistors (FETs). Usually, such devices incorporate solid back or top gate dielectrics. Here, we explore de-ionized (DI) water as a solution top gate for field-effect switching of layered semiconductors including $SnS_2$, $MoS_2$, and black phosphorus. The DI water gate is easily fabricated, can sustain rapid bias changes, and its efficient coupling to layered materials provides high on-off current ratios, a near-ideal sub-threshold swing, and enhanced short-channel behavior even for FETs with thick, bulk-like channels. Screening by the high-k solution gate eliminates hysteresis due to surface and interface trap states and substantially enhances the field-effect mobility. The onset of water electrolysis sets the ultimate limit to DI water gating at large negative gate bias. Measurements in this regime show promise for aqueous sensing, demonstrated here by the amperometric detection of glucose in aqueous solution. DI water gating of layered semiconductors can be harnessed in research on novel materials and devices, and it may with further development find broad applications in microelectronics and sensing.

**Significance Statement**

Layered semiconductors show promise as channel materials for field-effect transistors (FETs), but due to screening and limited field penetration the solid gate dielectrics usually employed in such devices can only control the conductance of ultrathin FET channels. Here we explore solution gating of layered FETs using de-ionized water as a gate dielectric. The DI water gate couples strongly to the channel, provides hysteresis free high on-off ratio switching, high carrier mobility and enhanced short-channel behavior even for FETs with thick, bulk-like channels. Water-gated FETs lend themselves naturally to aqueous sensing, demonstrated here by sensitive glucose detection. DI water gating of layered semiconductors can be harnessed in research on novel materials and devices, as well as broad applications in microelectronics and sensing.

**Introduction**

Field-effect transistors (FET), especially those made of silicon, play a central role in microelectronics (1, 2). Layered crystals, such as graphite, $MoS_2$, $SnS_2$, $WS_2$ and others whose electronic structure has long been known (3-10) have only recently been considered as channel materials for FET devices (11-14). One important reason is that although several layered semiconductors have sizable bandgaps, carrier transport in thick, bulk-like crystals could not be modulated effectively by electrical fields due to screening and the resulting limited penetration of the gate electric field. The successful isolation of single-layer graphene opened new possibilities for field-effect devices by demonstrating that the carrier density in atomically thin devices can be tuned by moderate applied electric fields, even in semimetals (15). Graphene FETs have shown very high carrier mobilities of up to $10^6$ $cm^2$/Vs, but the vanishing bandgap limits their application in digital logic because graphene transistors cannot be turned off, i.e., show large off-state currents (16). This limitation of graphene has led to efforts to exfoliate layered semiconductors, for example transition metal dichalcogenides such as $MoS_2$, to singlelayer thickness for field-effect devices and the exploration of the fundamental properties of such monolayer semiconductors (11, 13, 17, 18). The on-off ratio in $MoS_2$ FETs, for example, reaches up to $10^8$ at room temperature (11) and other 2D semiconductors ($WS_2$, $WSe_2$, $SnS_2$, etc.) showed similar values, as well as a host of other fascinating properties (19-25).

Previous studies of layered semiconductor FETs have mainly focused on single- and few-layer devices, with nearly all reports to date involving channel thicknesses below 10 layers (11, 12). The interest in the ultrathin limit is in part driven by emerging properties ue to the extreme carrier confinement in 2D crystals, but also reflects shortcomings of conventional gating strategies in controlling the conductance of thicker device channels. Back gating via a thick $SiO_2$ dielectric, the prevalent way of modulating electronic transport in studies of novel 2D and layered materials, offers relatively poor gate coupling and field penetration into layered crystals and hence cannot turn off devices with thicker, bulk-like channels. Top gates can achieve higher electric fields, but are more difficult to fabricate and still offer limited control over the conductance of thick FETs. Ionic liquids (26, 27) can be used to achieve very large electric fields in 2D and layered materials (28), sufficient for example to induce superconductivity in MoS$_2$ (29, 30). But ionic liquid gates face practical limitations. Slow charge transfer processes, for instance, require low bias scan rates to maintain equilibrium at the gate-channel interface. This calls for the development of alternative gating methodologies to study charge transport in layered materials and to harness the inherent advantages of FETs with thicker channels (31-34) compared to ultrathin single- or few-layer devices: higher current carrying capacity; lower Schottky barriers and hence reduced contact resistance at the layered semiconductor/metal junction (35, 36); and relative insensitivity to thickness fluctuations of the active layer, which facilitates robust, high-yield device fabrication over large areas.

Here, we explore the use of a de-ionized (DI) water top gate (13) for field-effect tuning of the conductance of thick layered-semiconductor FETs. This approach offers an alternative route toward efficient layered field-effect devices: instead of reducing the channel thickness to the ultimate 2D limit, the conductance of a thicker channel is switched between on- and off-states by a gate that supports very large electric fields at the channel surface. We find that the field penetrates sufficiently deep into the bulk that high on-off current ratios (up to $10^7$) can be obtained for device channels up to several hundred layers thick. In addition, the efficient gating implies that short-channel effects can be largely suppressed even for thicker channels. Since liquid gate drops are simply dispensed onto the semiconductor surface and contacted by a suitable electrode, the fabrication challenges of solid top-gate dielectrics are avoided. Gating via the double layer at the water-semiconductor interface supports much higher voltage sweep rates than ionic liquids. Finally, the use of an aqueous solution for gating provides facile access to water-based sensing, which we demonstrate here by detecting glucose with high sensitivity down to concentrations below 1 mM. While we focus mainly on FETs with layered tin disulfide (SnS$_2$, a n-type semiconductor with ~2.3 eV bandgap, (13)) channels to illustrate the characteristics of DI water top gates, the approach is generally applicable to a broader class of layered semiconductors with inert as well as more reactive surfaces, as demonstrated for devices with thick MoS$_2$ and black phosphorus channels.

**Results and Discussion**

Figure 1 illustrates the layout of the FET devices used here. The active channel consists of a layered semiconductor such as SnS$_2$, here prepared by exfoliation from highquality bulk crystals to a typical thickness of 50-60 nm (~80-100 layers) and supported on 300 nm $SiO_2$/Si. The contacts (Ti/Au, 5 nm/50 nm, see Methods for details) are covered by a thin polymer (poly-methyl methacrylate, PMMA) layer to electrically isolate them from the gate, which consists of a DI-water drop applied by pipette. The device can be gated alternatively by a top-gate voltage ($V_{TG}$) applied to an electrode in contact with the water drop, or by back gating ($V_{BG}$) via the $SiO_2$ support (Fig. 1a). Figs. 1 b-e show cross-sectional transmission electron microscopy (TEM) images of one of the fabricated devices (without the PMMA insulator layer). The active region, ~82 nm thick, is mostly homogeneous but shows a number of stacking faults and embedded smaller rains. In higher magnification TEM images, the layered structure is clearly visible (Fig. c, d) with periodicity of ~0.6 nm (Fig. 1e), consistent with the layer spacing of $SnS_2$ (13).

A direct comparison of the transfer characteristics of devices with back- and solution top gating demonstrates the striking effects due to the high electric fields at the water/layered metal dichalcogenide interface. The electric field applied via the $SiO_2$ back gate fails to bring the thick (~80 nm) $SnS_2$ channel into its off state, even for large applied voltages (-20 V, Fig. 2a). This is evidenced by a poor on-off current ratio (~10 over a wide range, $-20 \leq V_{BG} \leq 20$ V), as well as large "off-state" source-drain currents, $I_{SD}$ (Fig. 2b). The same device controlled by DI-water top gate shows substantially improved transconductance characteristics. $I_{SD}$ is now efficiently modulated via small changes in gate bias, $|V_{TG}| < 1$ V (Fig. 2c) and the channel is efficiently turned off, as demonstrated by on-off ratios exceeding $10^5$ (Fig. 2d). These values are sufficient for practical, low voltage operation of layered metal chalcogenide-based field-effect devices. Similar behavior is observed for other materials. Fig. 3 shows examples of device characteristics of FETs with thick $MoS_2$ and black phosphorus channels. As for $SnS_2$, a comparison of back-gating via the 300 nm $SiO_2$/Si support with top-gating by DI water shows significantly enhanced on-off ratios with the solution gate, even for channels with thickness of several tens of nanometers.

Top-gating by DI-water not only enhances the field penetration of thick device channels, allowing them to be turned off completely, but it also improves the current hysteresis during gate bias sweeps and results in a substantially higher carrier mobility. These improvements are illustrated in Fig. 4 for $SnS_2$ field-effect transistors. FETs fabricated from layered materials often show significant hysteresis, assigned primarily to charge trapping by surface and interfacial trap states (37). As can be seen in Fig. 4a, the transfer characteristic of back-gated SnS$_2$ devices indeed shows this hysteresis upon reversal of the gate bias sweep direction. The on-off ratio of this device is 67 over a gate voltage range from -35 to +40 V, and similar to the FET shown in Fig. 2 it cannot be turned off completely. When controlled by DI-water top gate, the same device shows almost no detectable hysteresis upon reversal of the gate voltage sweep direction (Fig. 4b), which implies that charge traps are no longer active after the water drop is applied. This behavior is consistent with the absence of surface adsorbates in the solution environment and an effective screening of interface states by the high-k dielectric ($\varepsilon_{rel}$(H$_2$O) ~ 80). Qualitatively similar results have also been reported for thin MoS$_2$ FETs employing HfO$_2$ or Al$_2$O$_3$ top gates (38-40). Again, solution top gating causes a striking enhancement of the on-off ratio to 5×10$^5$, nearly four orders larger than with the back gate and results in a near-ideal sub-threshold swing of 80 mV/decade (Fig. 4b). To compare the field-effect electron mobility for the two gate types, the mobility has been calculated from the transconductance characteristics via $\mu = \frac{dI_{SD}}{dV_G} \cdot \frac{L}{WCV_{SD}}$. The particular transistor considered in Fig. 4 had an aspect ratio $L/W = 3$, and the gate capacitance is either C(SiO$_2$) = 11.6 nF/cm$^2$ (back gate, (41)) or C(H$_2$O) = 137 nF/cm$^2$ (top gate), the double-layer capacitance of the DI water gate in contact with a SnS$_2$ FET channel determined previously (13). The field effect mobility increases from 50 cm$^2$/Vs (back gated) to 180 cm$^2$/Vs (top gated). This increase due to screening of scattering centers by the liquid dielectric is less pronounced than in the case of thinner few-layer or monolayer SnS$_2$ FETs since surface and interface scatterers play a relatively smaller role in the devices with thick channel considered here. Measurements of the on-off ratio on devices with different channel thickness (Fig. 4c) demonstrate that the overall gains due to liquid gating are nearly independent of the thickness of the active layer, so that on-off current ratios greater than 10$^5$ can be maintained for bulk-like channels up to at least 200 nm in thickness.

Layered semiconductors may offer superior scaling and improved short-channel characteristics compared with silicon FETs. We fabricated FETs with channel length ranging from 2 μm to 200 nm to compare the scaling of SnS$_2$ FETs with SiO$_2$ back gate and DI-water top gate (Fig. 5a). Devices with thin (~8 nm) and long channel show consistently high on-off ratios (>$10^6$), both controlled by back gating and solution top gating. However, for channel length below ~800 nm the on-off ratio of back-gated devices drops below $10^5$ whereas that of the top-gated devices remains high (Fig. 5b). A similar trend is observed for FETs with thick (55 nm) SnS$_2$ channel (Fig. 5c). The already low on-off ratio of back-gated devices drops even further for short channels with length below 1 μm. In contrast, the on-off ratio of devices controlled by DI water top gates remains high (~$10^6$) over the entire range of channel lengths considered here. This suggests that the very high electric fields achievable at the solid-liquid interface can effectively prevent possible leakage due to a finite source-drain bias at short channel lengths, so that the off-current remains low at least down to the shortest channels considered here (200 nm). This beneficial effect applies to both ultrathin devices, which have been considered extensively before (36, 42), as well as thicker, bulk-like layered semiconductor devices that are the focus of the present work. The observed trends can be understood within a framework developed for scaling in silicon, which invokes the characteristic length $\lambda = \sqrt{(\varepsilon_s/\varepsilon_{ox}) \cdot \tau_s \tau_{ox}}$, determined by the permittivities and thicknesses of semiconductor ($\varepsilon_s$, $\tau_s$) and gate dielectric ($\varepsilon_{ox}$, $\tau_{ox}$) (43). Within this picture, FETs with ultrathin channel have smaller $\lambda$, i.e., can be scaled to smaller dimensions. On the other hand, the thick devices considered here accentuate differences in short-channel effects between back- and solution top gating: reduced on-off ratio at short channel lengths for back gating, but consistently high on-off ratio with solution gating. The superior performance of the solution gate results from a combination of high permittivity ($\varepsilon$~ 80) and small thickness (~5-10 nm) of the double-layer gate. Our results therefore suggest that the combination of a solution top gate with a relatively inert layered semiconductor active region, which forms a stable interface with aqueous solutions, can allow harnessing the advantages of thicker FET channels while still permitting scaling to short channel lengths.

We now explore in more detail the properties of the DI-water top gate itself, in particular the possibility of current flow through the gate electrode (i.e., gate leakage). The gate current $I_G$ in top-gated SnS$_2$ devices was measured simultaneously with $I_{SD}$ as the top gate bias $V_G$ was swept for different applied source-drain voltages, $V_{SD}$. Representative results of these measurements are shown in Fig. 6a. The FET transconductance characteristics at different $V_{SD}$ reflect the same behavior observed for other devices (e.g., Fig. 1), i.e., efficient control of $I_{SD}$ by the solution gate. Whereas the device is clearly in its off state for gate bias between -0.2 V and -1.0 V, the current starts to increase again for larger negative $V_G$. The measurements shown in Fig. 6a indicate that this behavior is not due to ambipolar conduction that has been reported previously for some layered semiconductors, such as bulk and ultrathin $WSe_2$ (17, 44). For example, the magnitude of $I_{SD}$ is independent of source-drain bias, in contrast with the expected behavior for ambipolar FETs. Simultaneous measurements of the gate current show the same onset of increased current for $V_G \sim -1.0$ V, mirroring the behavior observed in $I_{SD}$. Hence, we attribute the apparent rise of $I_{SD}$ to a rapidly increasing current flow across the gate junction at large negative bias. Note that for positive $V_G$ there is consistently only a small amount of gate leakage, limited to ~10% of $I_{SD}$ for small $V_{SD}$ (20 mV) and below 3% of $I_{SD}$ for larger $V_{SD}$ (80 mV).

Further analysis (see below) shows that $I_G$ increases asymptotically to large values at an applied bias of $(-1.21 \pm 0.05)$ V, close to the standard potential for electrolysis of water (-1.23 V). This suggests that the onset of electrolysis sets an ultimate limit to the operating range of a DI water top gate; the $SnS_2$ cathode (the FET channel) appears to act as a remarkably efficient electrocatalyst that facilitates water electrolysis with minimal over potential. Highly reproducible characteristics of the miniature electrochemical cell in our DI water top-gated $SnS_2$ FETs (Fig. 6a) raise the possibility that the gate current could be used for aqueous sensing. In the simplest case it should be possible to detect the presence of ionic solutes via an increase in conductivity across the solution gate. Here, we demonstrate a less straightforward example of aqueous sensing, the measurement of the concentration of glucose – a non-ionic solute – in water.

Fig. 6b shows $I_G$-$V_G$ characteristics for different glucose concentrations [$C_6H_{12}O_6$] in the DI water gate drop. A stepwise increase in [$C_6H_{12}O_6$] causes well-defined changes in the $I_G$-$V_G$ curves close to the onset of electrolysis. To quantify these changes, the measured data were fitted by an empirical function $I_G = I_0/(V_G - V_0)$, which correctly represents the strongly increasing (i.e., essentially diverging) gate current at the electrolysis potential ($V_0$). The fit results are shown as lines in Fig. 6b and summarized in Fig. 6c. According to the analysis, the change in glucose concentration affects only the current amplitude ($I_0$) but not the asymptotic electrolysis potential ($V_0$), which remains constant for all values of [$C_6H_{12}O_6$] probed here (0–50 mM). $I_0$ rises logarithmically with increasing glucose concentration (Fig. 6c). This behavior can form the basis for a simple amperometric scheme to measure glucose concentration, which is detected via the gate current $I_G$ at any fixed gate bias near the electrolysis threshold ($V_G \leq$ -0.9 V). As shown in Fig. 6d, any choice of $V_{GSense}$ in this range will give the same logarithmic dependence of $I_G$ on [$C_6H_{12}O_6$], with a detection limit below 1 mM and high sensitivity in the concentration range corresponding to physiological glucose levels in human blood (~4 –6.5 mM). We note that although our data show the ability to sense glucose via an electrical signal in $SnS_2$-based FETs, additional work is needed to probe the specificity to a single solute in the presence of other species in an aqueous solution. The incorporation of additional enzymatic components, e.g., glucose oxidase used in conventional sensors (45), may be required to achieve specific sensing. Beyond the static gate solutions used here, one can foresee the integration of $SnS_2$ devices in microfluidic platforms for realtime electrical sensing of aqueous solutes.

**Conclusions**

We have demonstrated solution gating by deionized water as a method for achieve field effect control over the conductance of a wide range of layered semiconductors, including $SnS_2$, $MoS_2$, and black phosphorus. The efficient coupling and large electric field of the solution gate provides high on-off current ratios, a near-ideal sub-threshold swing, and enhanced short-channel behavior even for bulk-like channels with thickness of several 10 nm to well over 100 nm. Other advantages of the solution top gate include facile fabrication compared to conventional approaches involving solid dielectrics, and the capability of applying rapid changes in gate bias unlike ionic liquid gates that require slow bias ramps. Screening of surface and interface trap states by the high-k solution gate eliminates the gate-bias hysteresis found for back-gated devices and substantially enhances the field-effect mobility. These favorable characteristics of DI water gating in FETs can readily be leveraged in materials and device research, and they may find applications in microelectronics if suitable encapsulation strategies for a liquid gate can be identified. The onset of water electrolysis and associated leakage currents sets the ultimate limit to DI water gating at large negative gate bias. Measurements in this regime show promise for aqueous sensing, explored here by the amperometric detection of glucose in static drops down to concentrations of ~1 mM. By incorporating layered semiconductor FETs into microfluidic platforms, this approach could be expanded to allow high-throughput sensing in aqueous solutions.

**Acknowledgements**

This work was supported by the U.S. Department of Energy, Office of Science, Basic Energy Sciences, under Award No. DE-SC0016343. Device fabrication was carried out at the Center for Functional Nanomaterials, which is a U.S. DOE Office of Science Facility, at Brookhaven National Laboratory under Contract No. DE-SC0012704.

**Methods**
**1. Sample preparation.**

Adhesive tape was used to exfoliate layered bulk crystals ($SnS_2$, $MoS_2$, black phosphorus). It was folded 2-3 times to make the bulk crystal thinner. The tape with $SnS_2$ (or other layered material) was put onto the surface of a $SiO_2$/Si (300 nm) substrate cleaned by oxygen plasma. Before peeling off the adhesive tape, the substrate together with layered crystal/tape was put on a hot plate to anneal for 1-2 min at 100 °C to improve the contact area between the layered crystal and $SiO_2$ surface and obtain large flakes (46).

**2. Device fabrication and electrical measurements.**

After transfer of $SnS_2$ (or $MoS_2$, black phosphorus) flakes to $SiO_2$/Si substrate by mechanical exfoliation, optical microscopy was used to identify selected flakes with different thickness; the thickness was measured by AFM. Photoresist was spin-coated onto the substrates (S1811, 3000 rpm, 1 min), and annealed on a hot plate at 110°C for 2 min. Patterning was done using UV lithography (mask aligner Karl Suss, MA6), followed by resist developing in solvent. Ti/Au (5 nm/50 nm) contact metallization was performed in a lift-off process by e-beam evaporation, followed by photoresist dissolution in acetone. The devices were annealed in ultrahigh vacuum ($10^{-9}$ Torr) to enhance the contact resistance. Electrical measurements were performed on a four-probe station (Signatone). For DI water top-gated FETs, an additional PMMA layer was deposited on the devices, baked at 180°C for 2 min., and patterned by electron-beam lithography to open windows for contact between the water drops and the device channel.

**3. Transmission electron microscopy on complete devices.**

Fully functional FETdevices were covered in platinum and prepared into electron transparent cross-section specimens by focused ion beam (FIB) milling using a FEI Helios dual beam FIB instrument, followed by transfer to TEM grids. The FIB sections included the device channel and the adjacent source and drain contacts. TEM imaging was performed in a FEI Tecnai Osiris ChemiSTEM microscope at 200 kV beam energy.

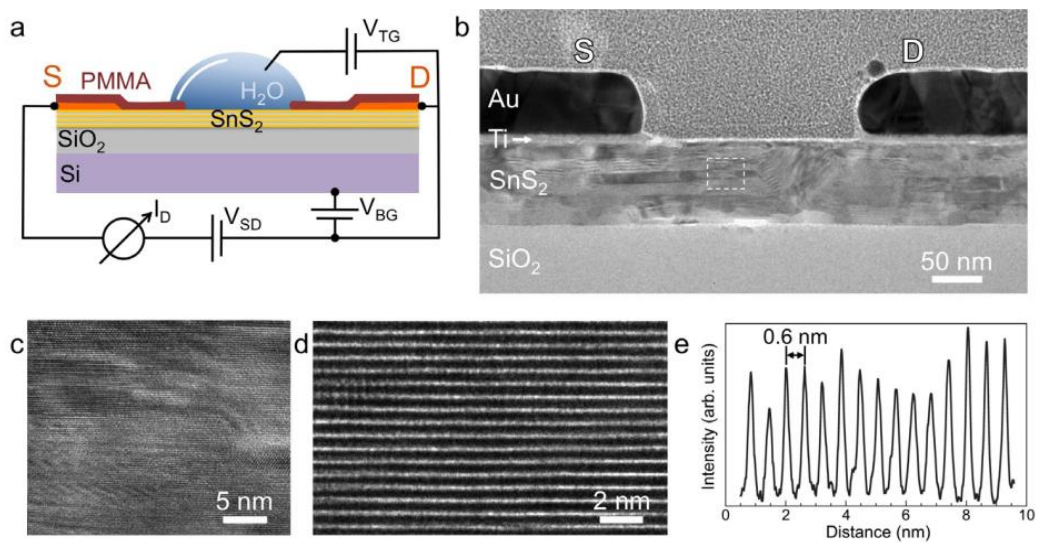

**Figure 1. Schematic and cross-sectional TEM of a field-effect transistor (FET) with multilayer SnS$_2$ channel. a.** Schematic geometry of layered metal chalcogenide semiconductor based devices (e.g., with SnS$_2$ channel). S: Source electrode; D: Drain; $V_{BG}$: Back-gate voltage, applied via the SiO$_2$/Si substrate; $V_{TG}$: Top-gate voltage, applied via a de-ionized water solution gate, insulated by PMMA from the source/drain contacts. **b.** TEM image of a FET device with 80 nm thick SnS$_2$ channel (~210 nm channel length) and Au/Ti contacts, supported on SiO$_2$/Si. **c.** High-magnification view of the region of the SnS$_2$ channel marked in a. **d.** High-resolution TEM image, showing the SnS$_2$ layering. **e.** Line profile showing a layer spacing of ~0.6 nm in the SnS$_2$ FET channel.

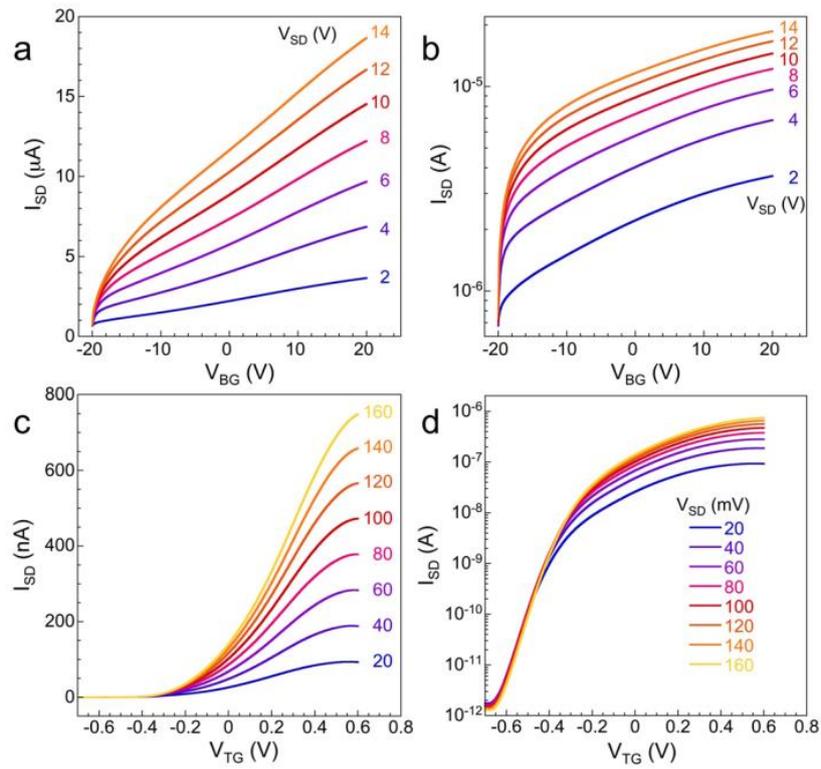

**Figure 2. Device characteristics of a thick (~80 nm) SnS$_2$ FET device. a.** Linear and **b.** logarithmic plots of the $I_{SD}$-$V_G$ transfer characteristics of the back-gated FET for source-drain bias $V_{SD}$ ranging from 2 to 14 V. For back-gate voltages between -20 - 20 V, the on-off ratio is less than 10. **c.** Linear and **d.** logarithmic plots of the $I_{SD}$-$V_G$ transfer characteristics of the solution top gated FET for source-drain bias curves at low bias $V_{SD}$ ranging from 20 to 160 mV. For top-gate voltages between -0.6 – 0.6 V the on-off ratio exceeds $10^5$.

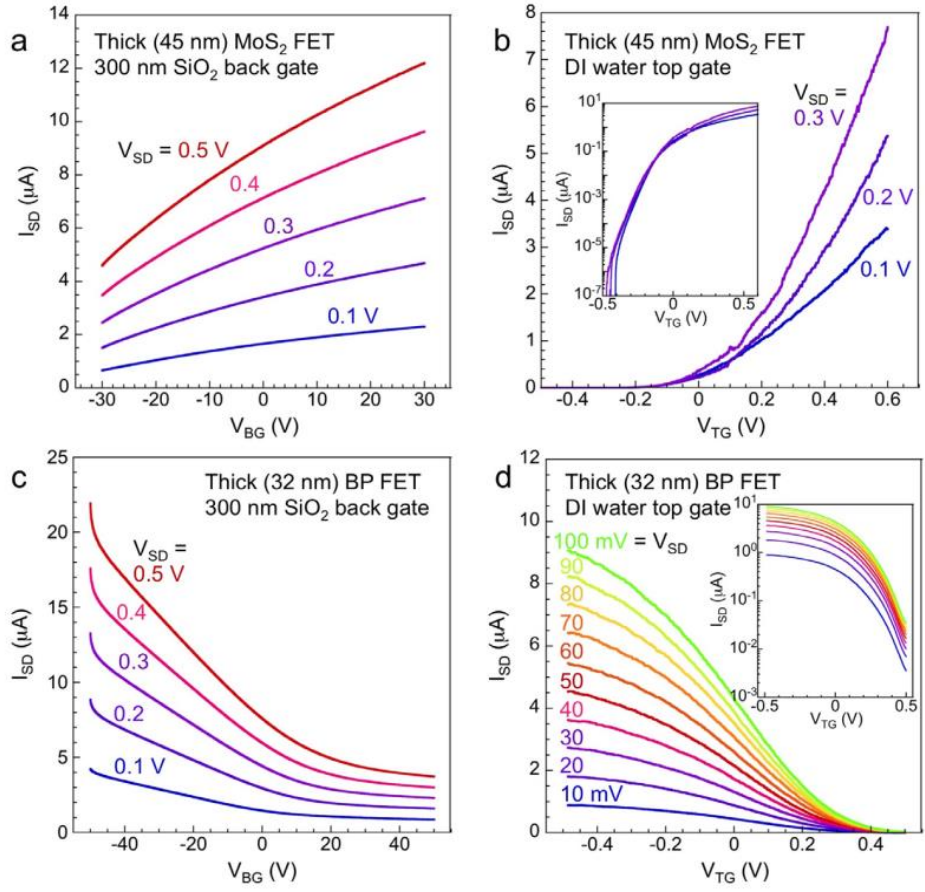

**Figure 3. Characteristics of thick MoS₂ and black phosphorus FETs. a.** $I_{SD}$-$V_G$ transfer characteristics of a back-gated MoS₂ FET with 45 nm channel thickness for source-drain bias $V_{SD}$ ranging from 0.1 V to 0.5 V. The back-gate is unable to bring the device into the off-state. **b.** Transfer characteristic of the same MoS₂ FET controlled by a DI water top gate. The logarithmic plot given in the inset shows an on-off ratio of ~$10^8$ of the top-gated device. **c.** $I_{SD}$-$V_G$ transfer characteristics of a back-gated black phosphorus (BP) FET with 32 nm channel thickness for source-drain bias $V_{SD}$ ranging from 0.1 V to 0.5 V. **d.** Transfer characteristic of the same black phosphorus FET controlled by a DI water top gate ($V_{SD}$ between 10 mV and 100 mV. Inset: logarithmic representation), showing reduced off-state current of the solution-gated device.

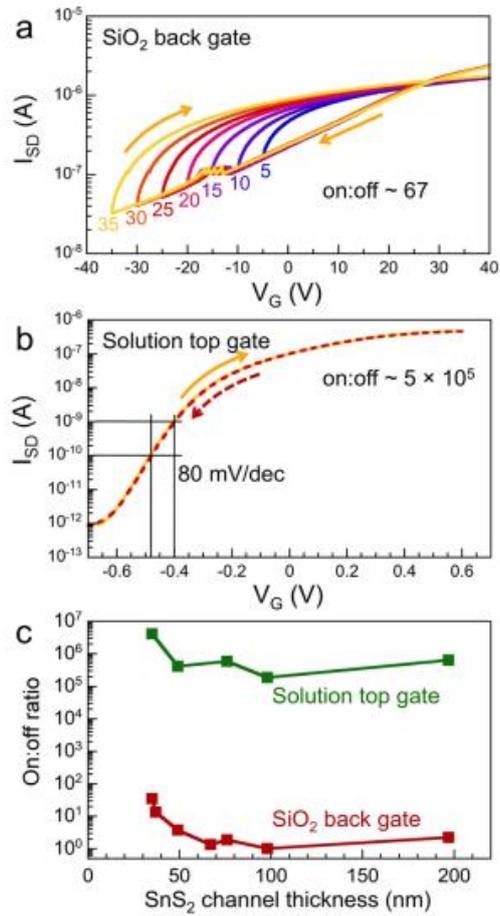

**Figure 4. Hysteresis and dependence of on:off current ratio on channel thickness for SnS$_2$ FETs. a.** Transfer characteristics of the back-gated FET, scanned from +40 V to negative gate voltages between -5 V and -35 V. Note the significant hysteresis as the V$_G$ sweep direction is reversed, and the small on:off current ratio. **b.** Transfer characteristic of the same device with water top-gating. Note the absence of hysteresis, large on:off current ratio (~5×10$^5$), and nearideal sub-threshold swing (80 mV/decade). **c.** Comparison of the on:off current ratios as a function of channel thickness for devices controlled by back gating and solution top gating.

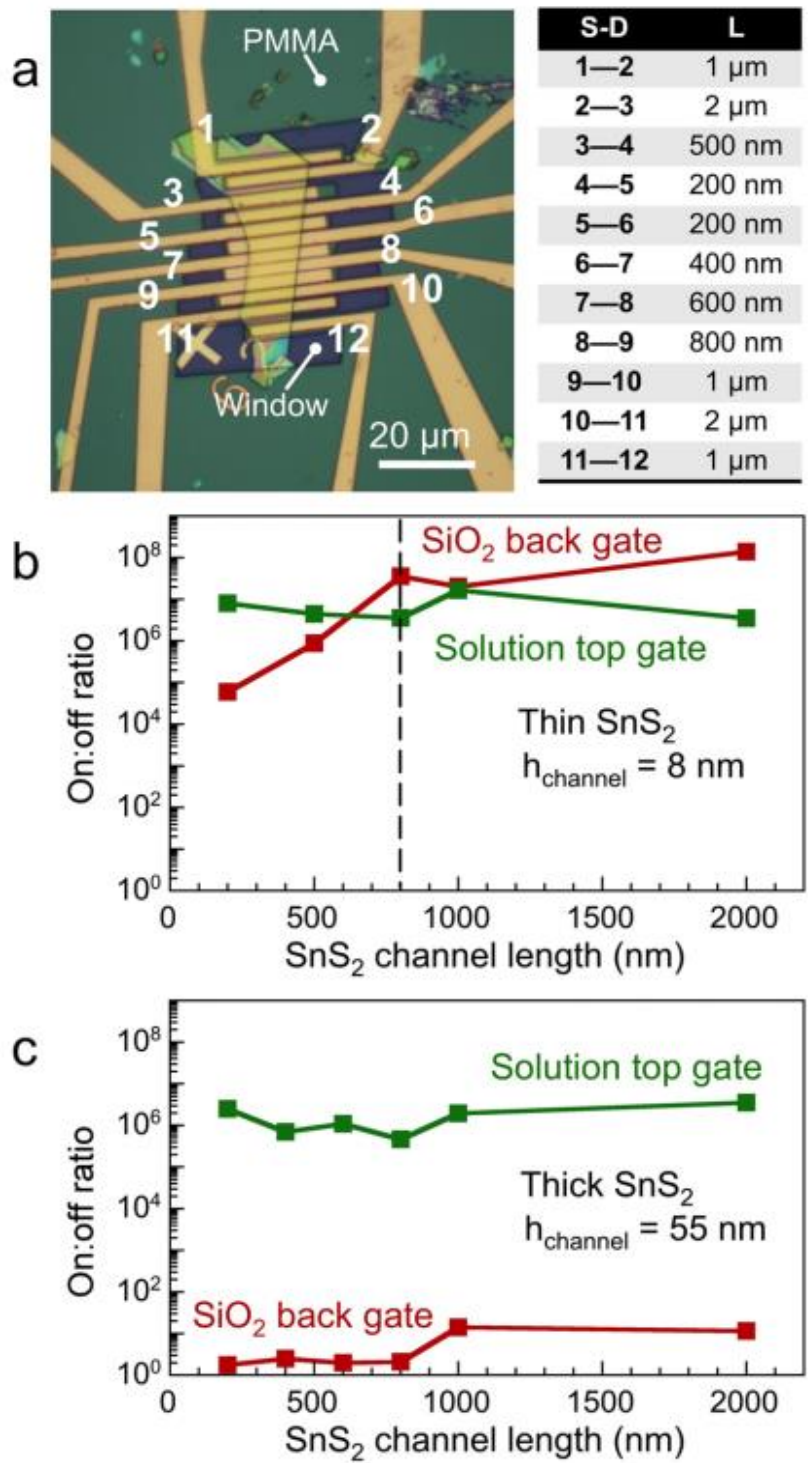

**Figure 5. Short-channel effects as reflected in the on:off ratio. a.** Device layout of SnS$_2$ FETs with different channel lengths ranging from 200 nm to 2 μm. **b.** Channel length scaling for SnS$_2$ devices with thin (8 nm, ~13 layers) channel. **c.** Channel length scaling for SnS$_2$ devices with thick (55 nm) channel.

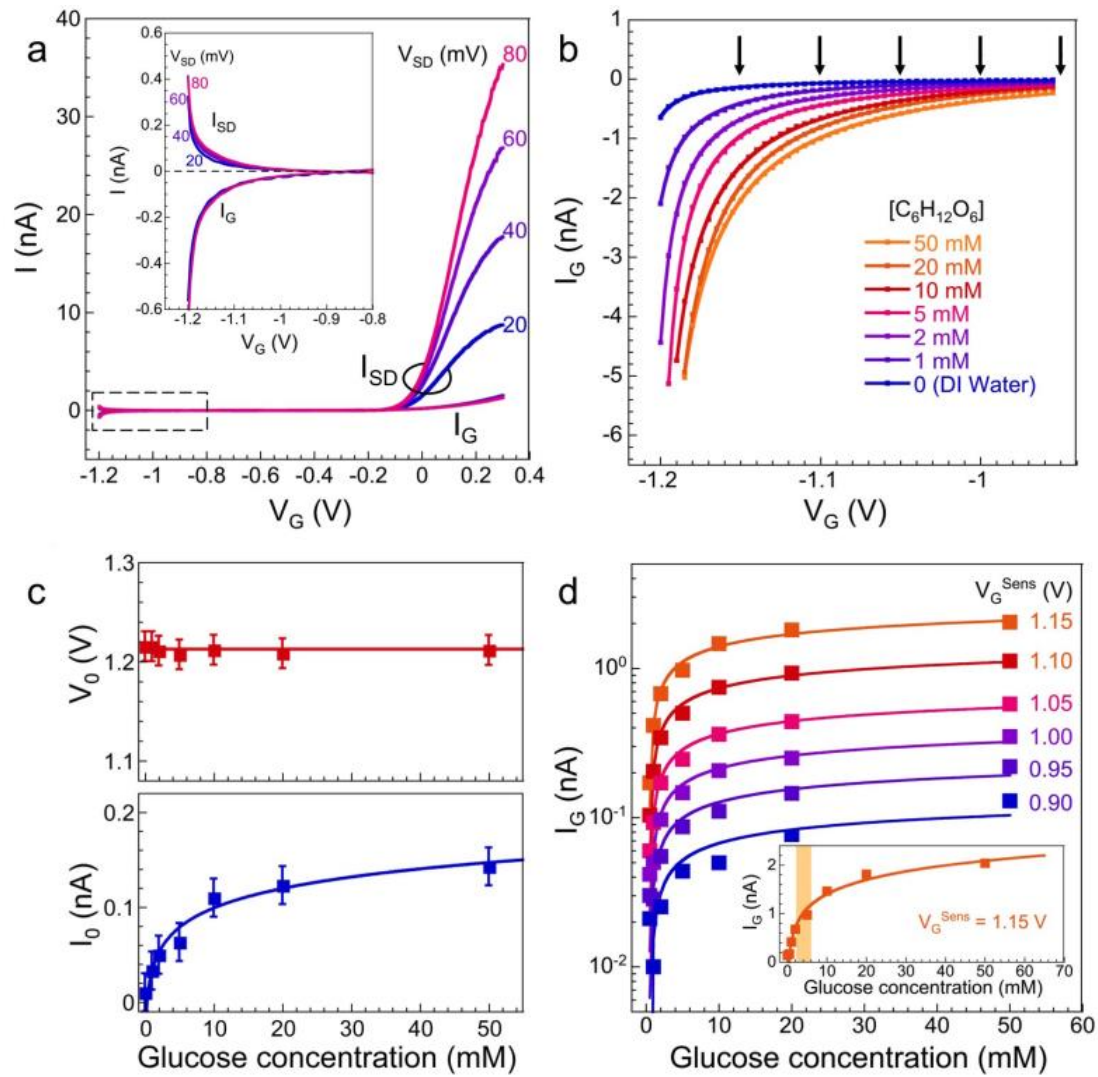

**Figure 6. Gate current and its use for aqueous sensing. a.** Source-drain current ($I_{SD}$) and gate current ($I_G$) of a FET with thick $SnS_2$ channel as a function of gate bias, applied to the DI water top gate. The zone of increasing $I_G$ at large negative gate voltage, marked by a dashed rectangle, is shown at higher magnification in the inset. **b.** Change in gate current with increasing concentration of glucose in the DI water gate. Lines are fits of $I_0/(V_G - V_0)$ to the data. **c.** Summary of the fit parameters derived from b, showing a constant asymptote voltage $V_0$ but logarithmically increasing current amplitude $I_0$ (solid blue line: log fit) with increasing glucose concentration. **d.** Glucose sensing via the gate current, $I_G$, at different fixed gate voltages (labeled; also indicated by arrows in b). Lines are logarithmic fits to the data. The inset indicates a detection limit below 1.0 mM glucose concentration.

18